\documentclass[twocolumn,amsmath,floats,showpacs,nofootinbib]{revtex4}

\usepackage{graphicx}

\usepackage{textcomp}
\usepackage{hyperref}

\hypersetup{colorlinks=true}

\begin{document}

\title{Reabsorption of nonequilibrium phonons at superconducting point contacts}

\author{I. K. Yanson, V. V. Fisun, N. L. Bobrov, and L. F. Rybal'chenko}

\affiliation
{B.I.~Verkin Institute for Low Temperature Physics and Engineering, of the National Academy of Sciences of Ukraine, prospekt Lenina, 47, Kharkov 61103, Ukraine\\
E-mail address: bobrov@ilt.kharkov.ua}
\published {\href{http://www.jetpletters.ac.ru/ps/141/article_2443.pdf}{Pis'ma Zh. Eksp. Teor. Fiz.} \textbf{45}, No. 9, 425 (1987); \href{http://www.jetpletters.ac.ru/ps/1244/article_18813.pdf}{JETP Lett.}, \textbf{45}, No. 9, 543 (1987)}
\date{\today}

\begin{abstract}When there is a deviation from the inequality $d\ll {{l}_{\varepsilon }}$ ($d$ is the contact diameter, and ${{l}_{\varepsilon }}$ is the energy relaxation length of the electrons), structural features are produced on the current-voltage characteristics of $S-c-N$ contacts at characteristic phonon energies because of the decrease in $\Delta $ due to the accumulation of nonequilibrium phonons with low group velocities near the contact.
\pacs{71.38.-k, 73.40.Jn, 74.25.Kc, 74.45.+c}
\end{abstract}
\maketitle

The nonlinear current-voltage characteristics of metal point contacts are known to undergo several changes in the region of characteristic phonon energies when one or both electrodes go superconducting. At clean $S-c-N$ contacts these changes stem from the emission of phonons in the course of an Andreev reflection of quasiparticles from the $N-S$ interface; at $\Delta \ll eV$, these changes usually amount to a small increment in the nonlinearity in the normal state \cite{Khlus,Yanson1}. In contacts with a potential barrier these changes stem from a tunneling component of the current. In the former case, the lines on the point-contact spectra of the electron-phonon interaction [on the curves of ${{d}^{2}}V/d{{I}^{2}}(V)$] are shifted to the left by a distance - $\Delta $ along the energy scale and are broadened by an amount on the order of the gap. In the latter case, the shift of the lines is $+\Delta $, and their shape is radically different from that of the spectral bands in the normal state \cite{Zakharov}.

We showed previously \cite{Yanson2} that sharp phonon structural features of unusual shape can be observed at $S-c-S$ point contacts with dimensions on the order of or greater than the energy relaxation length of the electrons, while in the normal state there are no spectral features of any sort (near the characteristic phonon frequencies). In the present study of $Ta-Cu$ and $Ta-Au$ heterocontacts with various resistances, we were able to trace a smooth transition from the point-contact spectra of $S-c-N$ contacts, which are similar to the ordinary spectra \cite{Khlus,Yanson1}, to the anomalous spectra described in Ref. \cite{Yanson2}. It has been found that the positions of the phonon features in the point-contact spectra of $S-c-N$ contacts of large dimensions $d\gtrsim \lambda (\varepsilon )$ correspond to the energies of phonons, $\hbar \omega $ (with low group velocities, $\partial \omega /\partial q\simeq 0$), which are slowly escaping from the vicinity of the contact and which locally reduce the energy gap. A new mechanism for the appearance of phonon structural features results from inelastic processes involving an Andreev reflection of electrons in the immediate vicinity of the $S-N$ interface. At bias energies $e{{V}_{s}}=\hbar {{\omega }_{s}}$, the excess current decreases along with  . These abrupt decreases in the excess current are manifested as spikes on the derivatives of the I-V characteristics.
\begin{figure}[]
\includegraphics[width=8.7cm,angle=0]{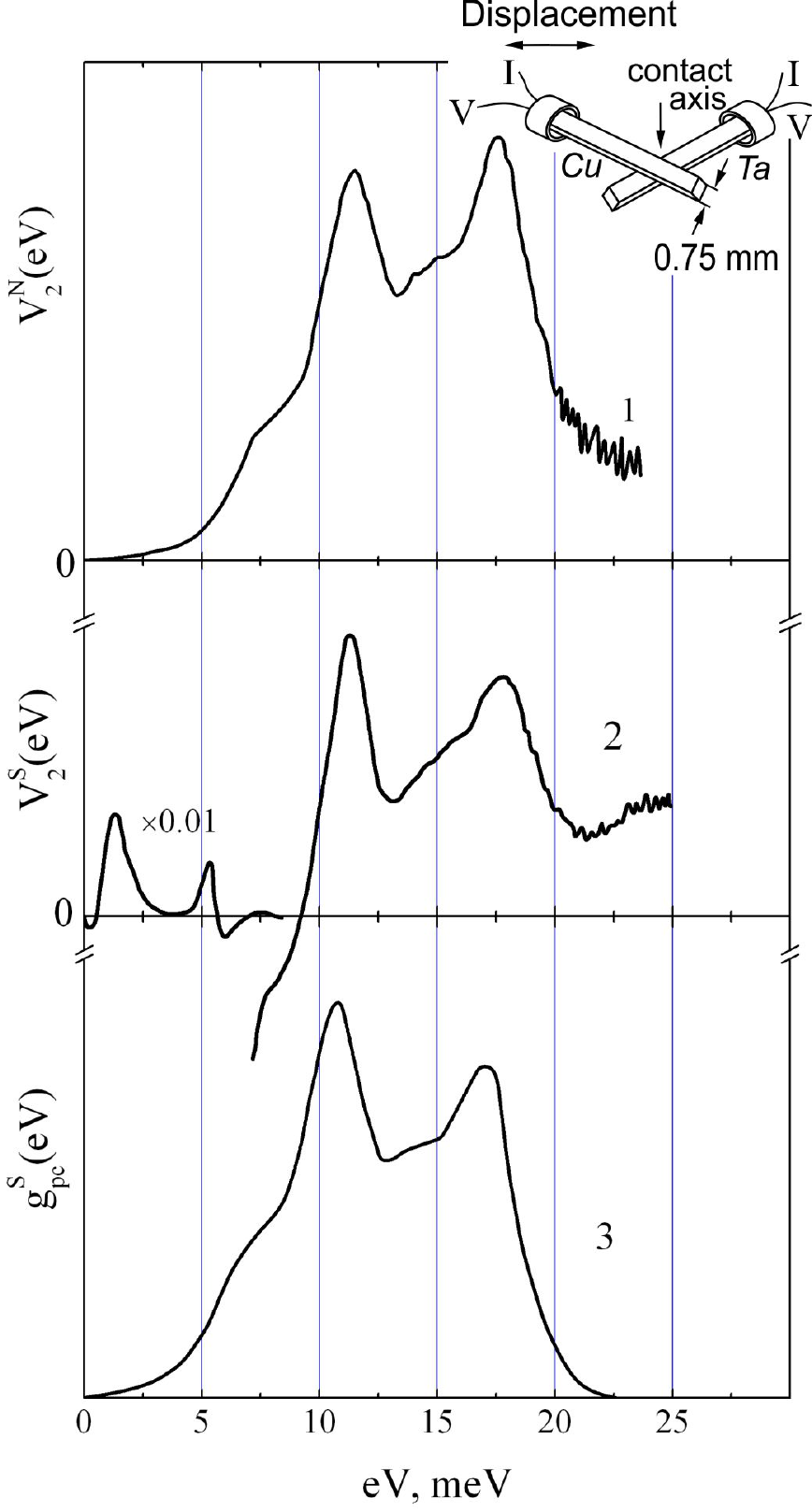}
\caption[]{\\1-Point-contact spectrum of the electron-phonon interaction of a $Ta-Cu$ heterocontact in the $N$ state [$R=80\ \Omega $, ${{V}_{1}}(0)=492\ \mu V$, $V_{2}^{\max }=0.767\ \mu V$, $T= 1.88 K$, $H=3 kOe$];\\
2 - spectrum of the same contact in the $S$ state [${{V}_{1}}(10)=502\ \mu V$, $V_{2}^{\max }=0.607\ \mu V$, $T=1.55 K$, $H = 0$; the part of the curve labeled $\times0.01$ corresponds to a reduction of the modulation by a factor of 10];\\
3 - calculated point-contact function of the electron-phonon interaction in the $S$ state found from spectrum 1 by subtracting the background. The inset shows the experimental geometry.}
\label{Fig1}
\end{figure}

Figure \ref{Fig1} shows the experimental geometry (in the inset) and point-contact spectra of $Ta-Cu$ in the normal and superconducting states (curves 1 and 2). The contact diameter can be estimated from ${{d}_{Ta-Cu}}\simeq {70}/{\sqrt{{{R}_{0}}}}\ (nm)$ (Ref. \cite{Bobrov}). It turns out to be 78 \AA\ although this figure is comparable to the electron-phonon scattering length at $eV\sim \hbar {{\omega }_{0}}$ in $Ta$, it is nevertheless slightly smaller [${{l}_{\varepsilon }}\simeq {{v}_{F}}{{\tau }_{\varepsilon }}\sim $120 \AA, where ${{v}_{F}}=0.51\times {{10}^{8}}\ cm/s$ and ${{\tau }_{\varepsilon }}=(2\pi /\hbar )\int\limits_{0}^{\varepsilon}{g(\omega )d\omega }$]. Subtracting the background from curve 1, we find the point-contact function of the electron-phonon interaction, $G(\omega )$, which we will use to calculate the spectrum in the superconducting state \cite{Khlus}:
\[\frac{1}{R}\frac{dR}{dV}=\frac{16ed}{3\hbar {{v}_{F}}}\int\limits_{0}^{\infty }{\frac{d\omega }{\Delta }S\left( \frac{\hbar \omega -eV}{\Delta } \right){{G}_{\alpha }}(\omega )\quad \quad T\ll \Delta }\]
where
\[S(x)=\Theta (x-1)\frac{2{{\left( x-\sqrt{{{x}^{2}}-1} \right)}^{2}}}{\sqrt{{{x}^{2}}-1}}\]

(curve 3 in Fig. \ref{Fig1}). Even for such a comparatively high-resistance contact there are some obvious differences between the observed spectrum (curve 2) and that which was expected (curve 3). The lines are not shifted by - $\Delta $, and the $T$-phonon band contracts significantly in the $S$ state instead of broadening. On spectrum 2 there is an intense structural feature (note the change in the ordinate scale) at $eV\sim 6\ meV$. This feature, which corresponds to an abrupt decrease in the excess current on the I-V characteristic, is caused by a decrease in $\Delta $ by an amount on the order of a few percent when the critical quasiparticle injection power ${{P}_{k}}$ is reached. The position of this structural feature along the $V$ scale corresponds to $V_{k}^{2}/R={{P}_{k}}=\text{const}$ and depends on the resistance of the contact, the magnetic field, and the temperatures \cite{Yanson3}. Interestingly, at clean contacts the self-magnetic field of the current near the microbridge, $H\simeq 4I/cd$, is proportional to the power $P=I\cdot V$, so that the critical power corresponds to a completely definite strength of the axisymmetric magnetic field near the contact. For $Ta-Cu$ contacts at $T = 2 K$ we find ${{P}_{k}}=0.46\ \mu \text{W}$ and ${{H}_{k}}=70\text{ Oe}$ - a fairly high figure. Although pure $Ta$ is a type I superconductor, even a very slight contamination is capable of converting it into a type II superconductor. It is extremely likely that the injection of nonequilibrium quasiparticles increases the Ginzburg-Landau parameter $k$ thus ${{H}_{k}}$ could definitely correspond to ${H}_{c1}$, and the abrupt decrease in the excess current at $V={{V}_{k}}$ might be due to a penetration of toroidal vortices into the contact region of the nonequilibrium superconductor.

\begin{figure}[]
\includegraphics[width=8.5cm,angle=0]{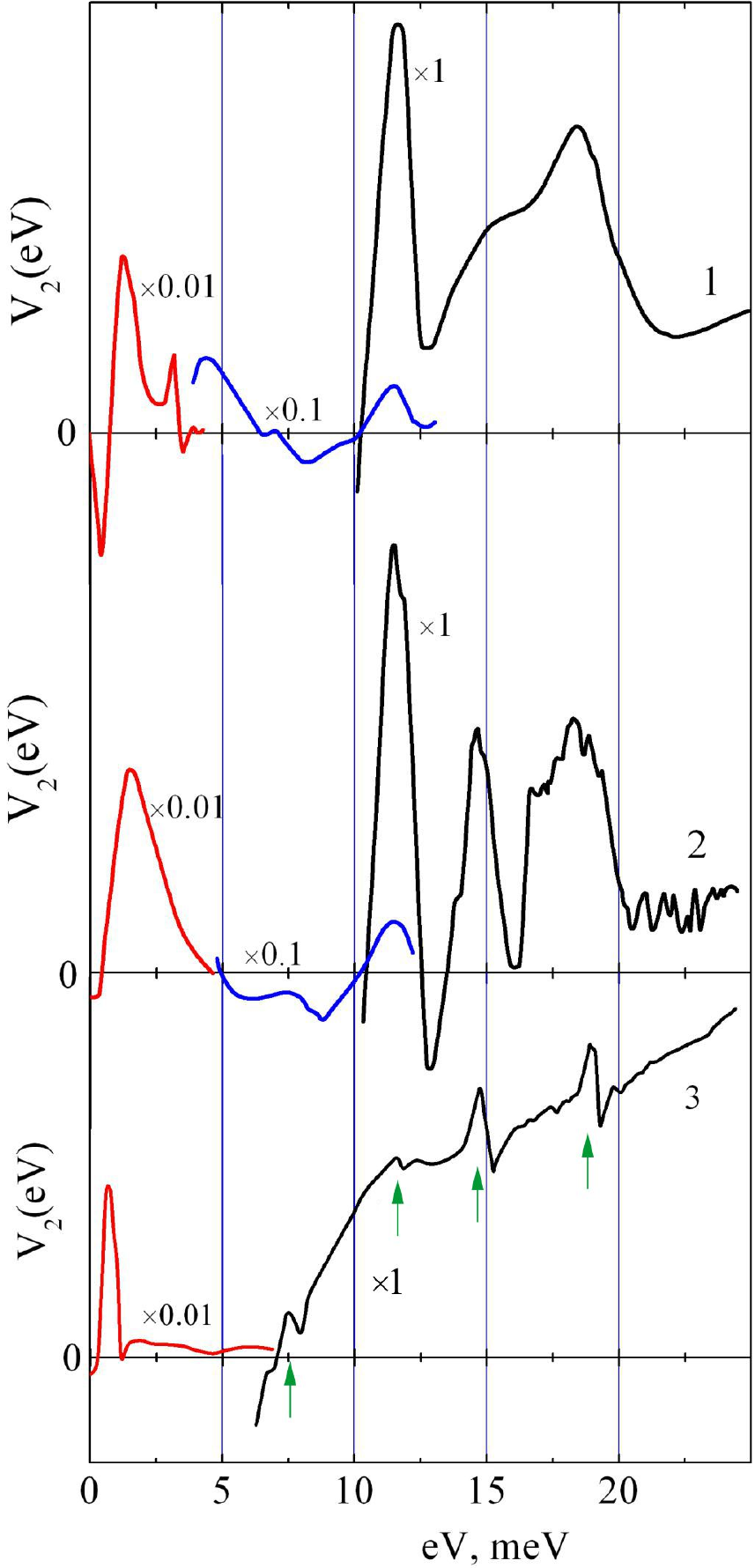}
\caption[]{\\Point-contact spectra of the electron-phonon interaction of heterocontacts in the $S$ state. $H=0$. The parts of the curves labeled $\times 0.1$ and $\times 0.01$ correspond to reductions of the modulating voltage by factors of $\sqrt{10}$ and 10, respectively.\\
1 - $Ta-Cu$, $T= 1.63 K$, $R=26.5\ \Omega $, ${{V}_{1}}(10)=448\ \mu V$, $V_{2}^{\max }=0.66\ \mu V$;\\
2 - $Ta-Au$, $T= 1.45 K$, $R=19\ \Omega $, ${{V}_{1}}(10)=829\ \mu V$, $V_{2}^{\max }=0.7\ \mu V$;\\
3 - $Ta-Au$, $T= 1.6 K$, $R=0.76\ \Omega $, ${{V}_{1}}(10)=320\ \mu V$, $V_{2}^{\max }=0.5\ \mu V$.\\
Here ${{V}_{1}}(10)$ is the effective value of the modulating voltage at a bias voltage of $10 mV$, and $V_{2}^{\max }$ is the effective value of the second harmonic at the absolute maximum of the spectrum.}
\label{Fig2}
\end{figure}

Let us examine the change in the shape of the phonon peaks on the point-contact spectra of $S-c-N$ contacts as the resistance is reduced, and the inequality $d\ll {{\Lambda }_{\varepsilon }}(eV)$ is violated. On the spectra we see a pronounced sharpening of the lines. The structural features at $eV\sim \text{7}$ and $15\ meV$, which are seen as no more than knees in the normal state and on the spectra of high-resistance contacts, are now clearly defined lines having the shape of a sine-wave cycle. The spectra of the $S-c-N$ contact of $Ta-Cu$ and $Ta-Au$ of intermediate resistance ($\text{3}-\text{3}0\ \Omega $) have essentially the same shape (curve 1 in Fig. \ref{Fig2}). Spectrum 2 corresponds to a diffusion regime (a short momentum mean free path of the electrons), since it has an absolute intensity several times lower.
\begin{figure}[]
\includegraphics[width=5cm,angle=0]{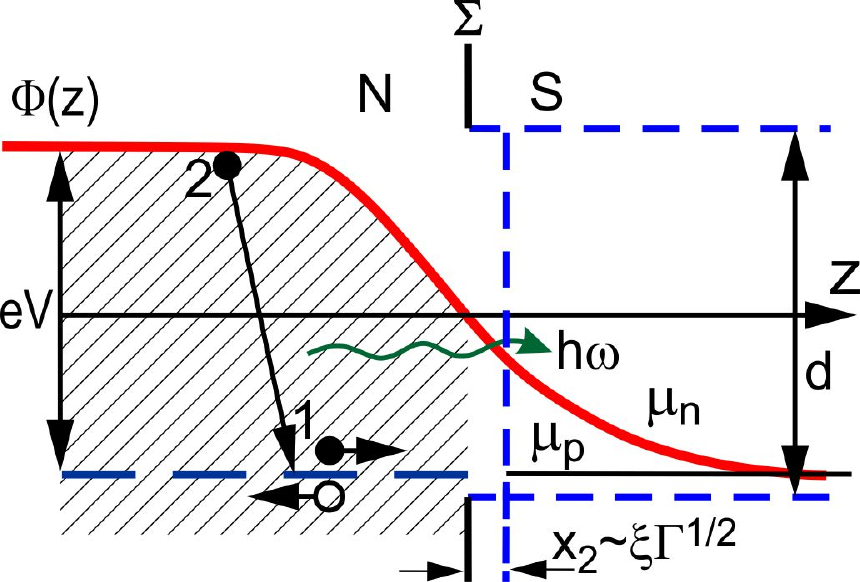}
\caption[]{Energy diagram of an $S-c-N$ contact. Here $\Sigma $ is an opaque screen; $\xi$ is the coherence length; $\Gamma \simeq 0.16(\hbar /{{\tau }_{ph}}{{k}_{B}}{{T}_{c}})\simeq 5\times {{10}^{-3}}$ is the depairing factor \cite{Ivlev}; ${{\mu }_{n}}$ is the electrochemical potential of the quasiparticles; and ${{\mu }_{p}}$ is the electrochemical potential of the pairs. The vertical arrow marks the inelastic transition of an electron 2 to a free state which arises during Andreev reflection of electron 1. A phonon with an energy $\hbar \omega =eV$ is emitted.}
\label{Fig3}
\end{figure}

It was found possible to fabricate extremely low-resistance contacts ($R<1\ \Omega $) for the $Ta-Au$ system (curve 3 in Fig. \ref{Fig2}). The characteristics of these contacts have little in common with the electron-phonon-interaction functions (curves 1 and 3 in Fig. \ref{Fig1}). Nevertheless, the positions of the sharp structural features against the monotonically increasing background (marked by the arrows) correspond to the energies of phonons with low group velocities in $Ta$ (according to Ref. \cite{Woods}, $\hbar {{\omega }_{s}}=$7.0, 11.3, 15.5, and 18 $meV$). In the normal state, there are no spectral features of any sort on the I-V characteristics of such contacts.

While the positions of the major $T$ and $L$ peaks on the point-contact spectra for the normal state vary over the respective intervals 11.5-12.5 $meV$ and 17-18 $meV$ for various contacts, in the superconducting state the positions of these peaks are fixed at bias energies of 11.3 and 18 $meV$, within an error of $\pm 0.1\ meV$. The stabilization of the positions of the peaks along the $eV$ axis at the transition to the superconducting state become progressively more obvious as the deviation from the condition for a ballistic situation, $d\ll {{l}_{\varepsilon }}$, increases (i.e., for low-resistance or dirty contacts).

The effect described here can be explained in terms of a reabsorption of nonequilibrium phonons which are generated in the immediate vicinity of the plane of the contact, where electrons which "remember" their energy far from the aperture are encountered (Fig. \ref{Fig3}). In the pure normal metal ($Cu$ or $Au$) the condition $d\ll {{l}_{\varepsilon }}$ holds. Although this condition does not hold for normal excitations in a superconductor, the condensate electrons reach the interface, holding the electrochemical potential constant over space out to distances from the interface on the order of the inelastic electron-phonon relaxation lengths or even less, as at phase-slippage centers.\footnote{The diameter of the contact must nevertheless be shorter than the depth to which the electric field penetrates into the superconductor, $\sim \xi /{{\Gamma }^{{1}/{2}\;}}$.} Accordingly, phonons with a maximum energy $\hbar \omega =eV$ are created near the interface. The elastic scattering of phonons hinders their escape and sharpens the selection effect in terms of the characteristic $\partial \omega /\partial q\simeq 0$. The reabsorption of nonequilibrium phonons in the superconductor leads to a decrease in $\Delta $ near the interface and therefore a decrease in the excess current. There is no shift of the spectral lines by an amount - $\Delta $ here. The mechanism proposed here obviously also operates at $S-c-S$ contact with dimensions greater than the energy relaxation length of the electrons and greater than the electrodynamic coherence length \cite{Yanson2} $\xi^{-1} =\left( \xi _{0}^{-1}+{{l}^{-1}} \right)$.

In the normal state, the reabsorption of phonons gives rise to a background on the point-contact spectra. The experimental data provide evidence that the spectrum in the $S$ state becomes progressively more different from that predicted by the theory of Ref. \cite{Khlus} as the background increases. This circumstance might also be regarded as indirect evidence for a role of nonequilibrium phonons in the changes in the shape of the point-contact spectrum at the transition to the $S$ state.

\end{document}